\documentclass[letter]{aa} 

\usepackage{graphicx}
\usepackage{txfonts}
\usepackage{natbib}
\usepackage{url}

\begin{document}

   \title{Hot Stuff for One Year (HSOY)}

   \subtitle{A 580 million star proper motion catalog derived from Gaia DR1 and PPMXL}

   \author{M. Altmann
          \inst{1}
          \and
          S. Roeser
          \inst{1,2}
          \and
          M. Demleitner
          \inst{1}
          \and 
          U. Bastian
          \inst{1}
          \and
          E. Schilbach
          \inst{1,2}
          }

   \institute{Zentrum f\"ur Astronomie der Universit\"at Heidelberg,
              Astronomisches Rechen-Institut, 
              M\"onchhofstr. 12-14,
              69120 Heidelberg, Germany
              \email{maltmann@ari.uni-heidelberg.de}
           \and
              Zentrum f\"ur Astronomie der Universit\"at Heidelberg,
              Landessternwarte K\"onigsstuhl,
              K\"onigsstuhl 12,
              69117 Heidelberg, Germany
            }

   \date{Received January 04, 2017, accepted xxx}

 
  \abstract
   {Recently, the first installment of data from ESA's Gaia astrometric satellite mission (Gaia DR1) was released, containing positions of more than 1 billion stars with
unprecedented precision, as well as proper motions and parallaxes, however only for a subset of 2 million objects. The second release
will include those quantities for most objects.} 
   {In order to provide a dataset that bridges the time gap between the Gaia DR1 and Gaia DR2 releases and partly remedies the lack of proper motions in the former,
 HSOY (''{\bf H}ot {\bf S}tuff for {\bf O}ne {\bf Y}ear'')
was created as a hybrid catalog between Gaia and ground-based astrometry, featuring proper motions (but no parallaxes) for a large fraction of the DR1 objects.
While not attempting to compete with future Gaia releases in terms of data quality or number of objects, the aim of HSOY is to provide improved proper motions partly
based on Gaia data, allowing studies to be carried out just now or as pilot studies for later projects requiring higher-precision data.}
   {The HSOY catalog was compiled using the positions taken from Gaia DR1 combined with the input data from the PPMXL catalog, employing the same
    weighted least-squares technique that was used to assemble the PPMXL catalog itself.}
   {This effort resulted in a four-parameter astrometric catalog containing 583 million stars, with Gaia DR1 quality positions and proper motions with precisions from
 far less than 1 mas/yr to 5 mas/yr, depending on object brightness and location on the sky.}
   {}


   \maketitle
%
\section{Introduction}
\label{sect:intro}
Gaia, ESA's 1-billion star astrometric space mission \citep{Gaia1} set out in late 2013 to revolutionise (among many other fields of astronomy) our understanding of
the kinematics, the structure and evolution of our Galaxy. On September 14, 2016, the Gaia consortium published the first Gaia Data release, Gaia DR1
\citep{Gaia2}, containing the positions and broad-band photometry for 1.143 billion objects. 
\begin{figure}
   \centering
   \includegraphics[width=\columnwidth]{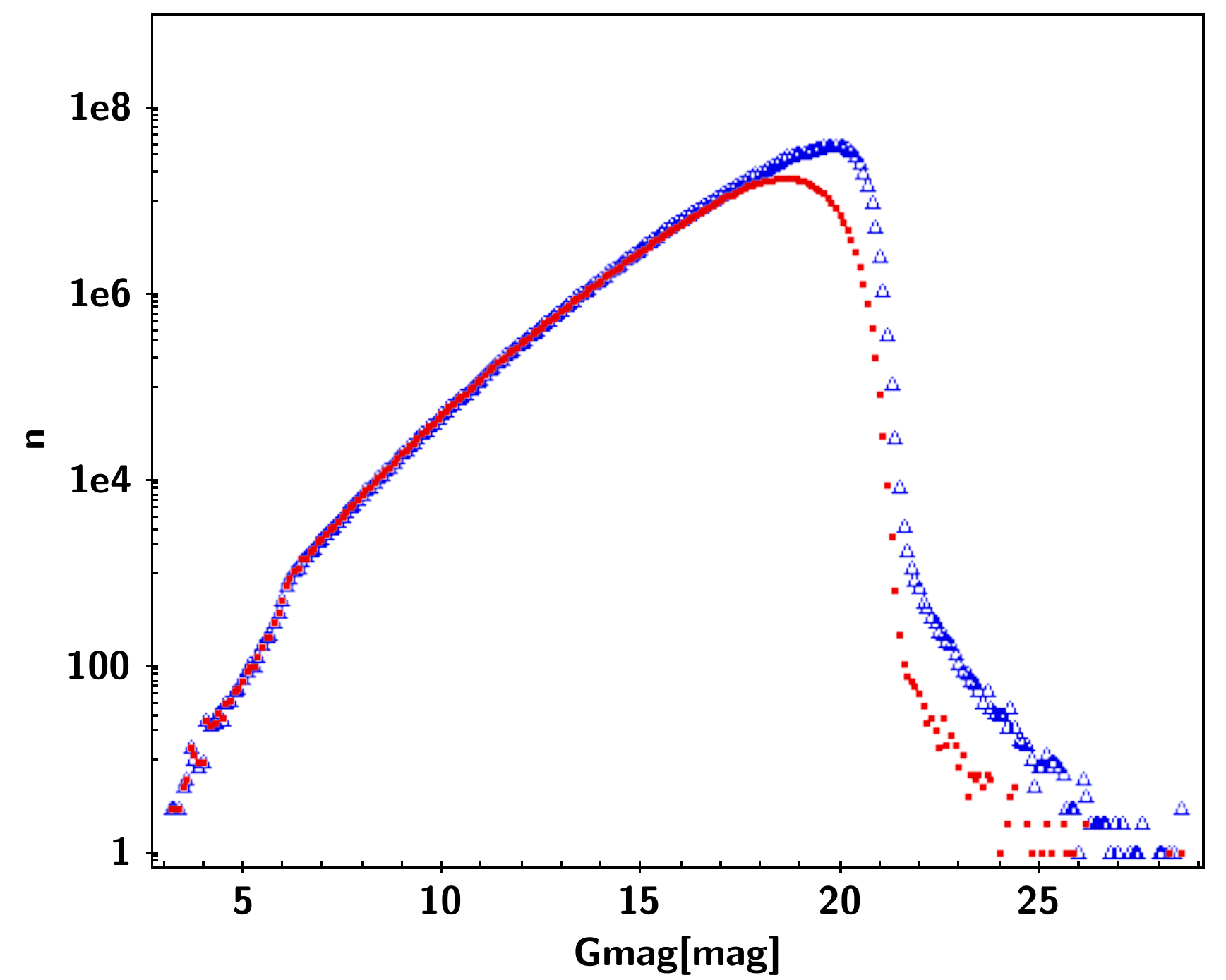}
   \caption{Distribution of object counts in HSOY over the Gaia $G$ magnitude(red filled circles). As comparison, the object counts for Gaia DR1 are also shown (blue open triangles).}
              \label{hsoy_logn.fig}%
    \end{figure}
   \begin{figure}
   \centering
   \includegraphics[width=\columnwidth]{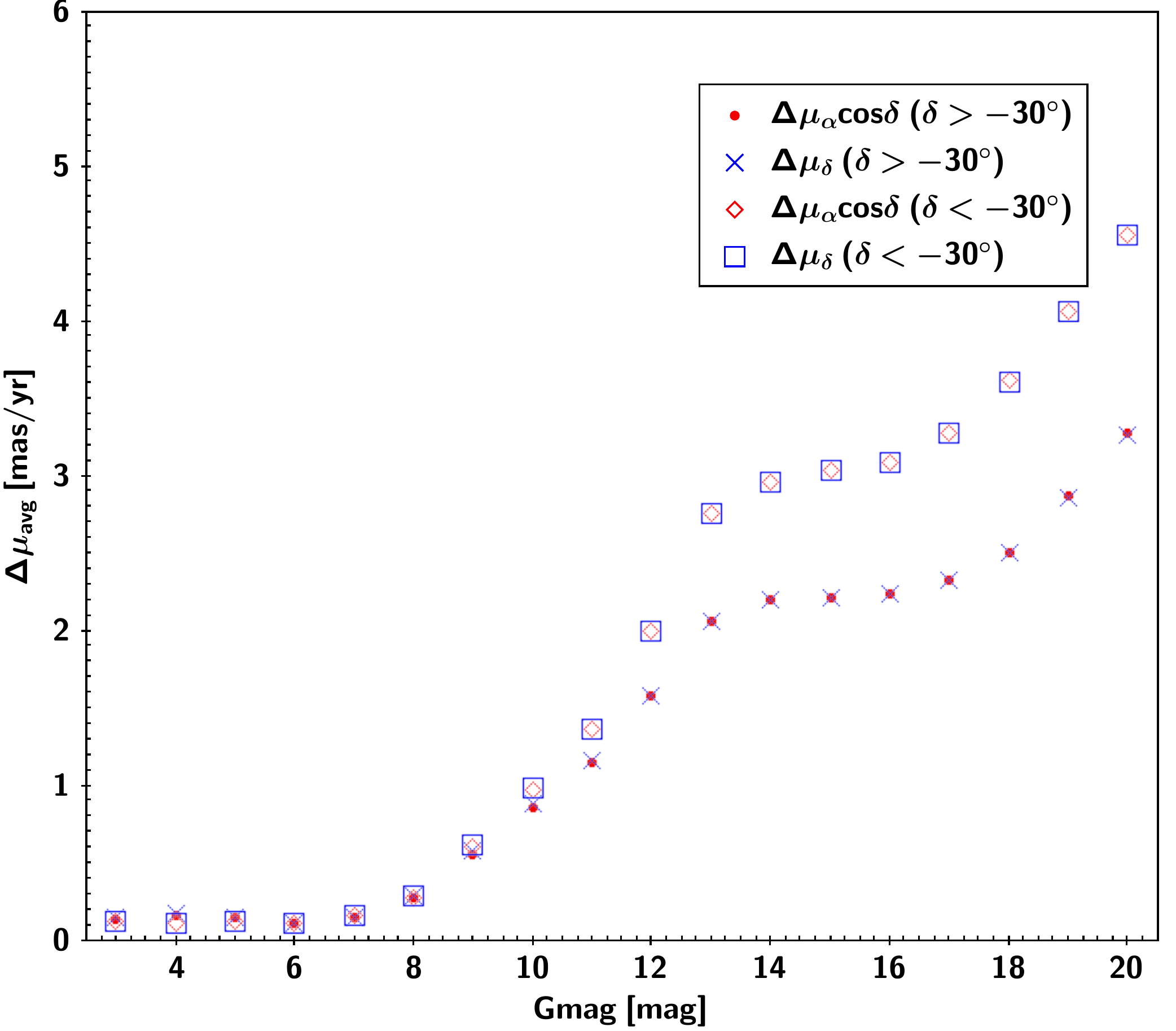}
   \caption{Mean standard errors in proper motions of the HSOY catalog north and south of $\delta=-30\degr$. The errors are given in bins of one magnitude centered on full Gaia $G$ magnitudes. The symbols are defined in the legend box within the plot}
              \label{hsoy_pmerrs.fig}%
    \end{figure}
Given the short timespan of the measurements incorporated into this release (less than 15 months), the separation of parallaxes and proper motions was not possible. 
Therefore Gaia DR1 in general does not include these quantities, with the exception of a small subset of those 2 million stars already present in the Hipparcos \citep{1997ESASP1200.....E}
or Tycho2 catalogs \citep{2000A&A...355L..27H}. Using this data as a first epoch, proper motions and parallaxes  could be disentangled, yielding the 
Tycho Gaia Astrometric Solution \citep[TGAS,][] {2015A&A...574A.115M}.
Thus most astrometry-related 
projects within the astronomical community might focus on the TGAS data, as well as on preparing for the next Gaia release.

   \begin{figure}
   \centering
   \includegraphics[width=\columnwidth]{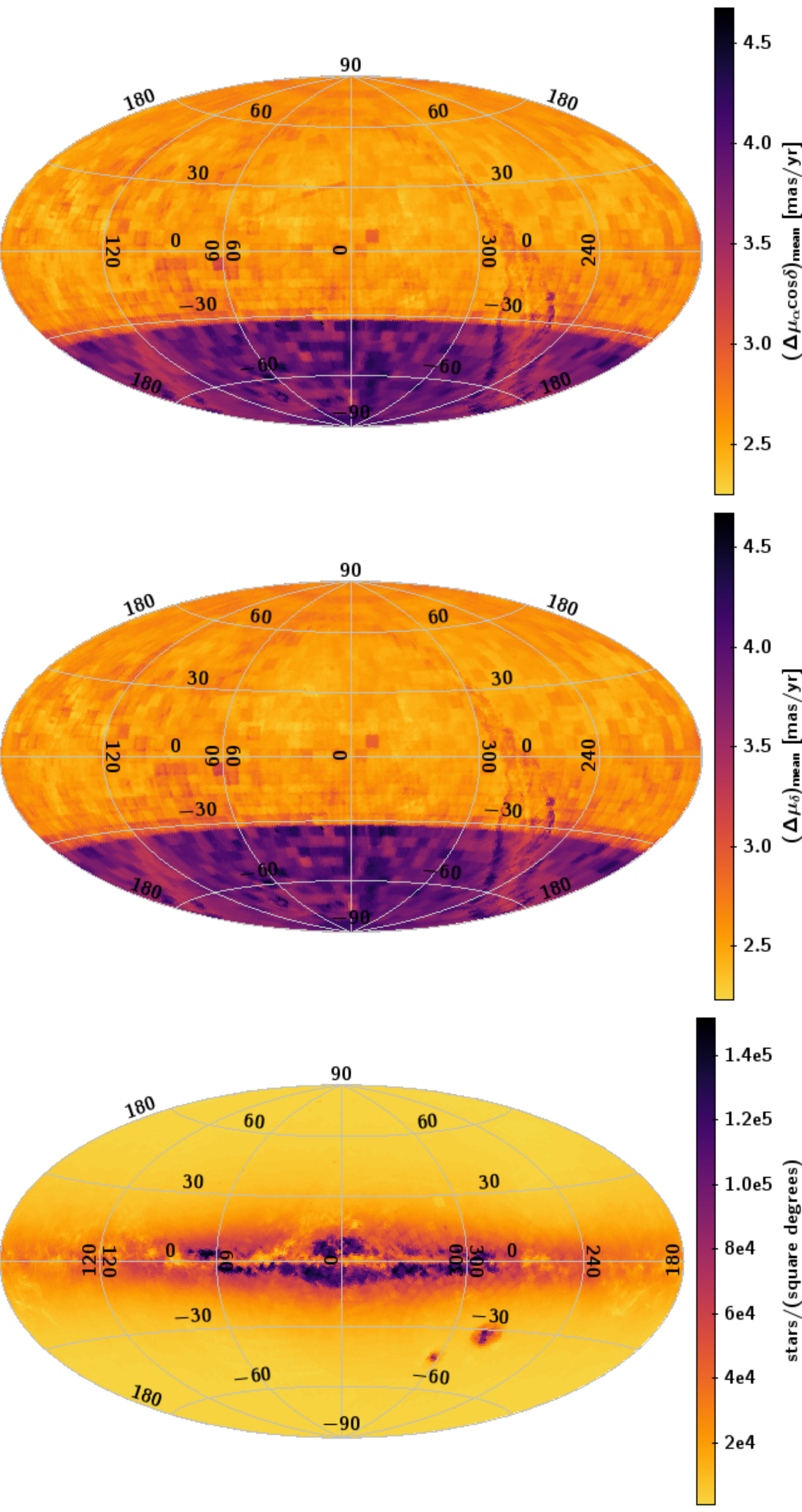}
   \caption{The upper two panels show the global variation of the r.m.s. errors in proper motion (upper panel, right ascension, middle panel, declination) in equatorial
coordinates. Darker shades/colours
indicate higher errors. 
 The bottom panel shows the density of HSOY in stars per square degrees over the entire sky in galactic coordinates
highlighting the overall uniformity of the catalog.}
              \label{hsoy_global.fig}%
    \end{figure}
\begin{table}
\caption{Content of the HSOY catalog. }
\label{tab1.tab}      
\centering                                      
\begin{tabular}{ r r r p{4.8cm}}          
\hline\hline                        
 &  Name & Unit & description \\
\hline
1 & $ipix$ & -- &  PPMXL object identifier \\
2 & $comp$ & -- & flag indicating multiple Gaia matches to one PPMXL object\\
3 & $raj2000$ & degrees & RA at J2000.0, epoch 2000.0\\
4 & $dej2000$ & degrees & Decl. J2000.0, epoch 2000.0\\
5 & $e\_ra$ & degrees & Mean error: $\alpha\cos\delta$ at mean epoch \\
6 & $e\_de$ & degrees & Mean error: $\delta$ at mean epoch \\
7 & $pmra$ & deg/yr & Proper motion in $\alpha\cos\delta$\\
8 & $pmde$ & deg/yr & Proper motion in $\delta$\\
9 & $e\_pmra$ & deg/yr & Mean error in $\mu_\alpha\cos\delta$\\
10& $e\_pmde$ & deg/yr & Mean error in $\mu_\delta$\\
11& $epra$ & yr & Mean Epoch in RA ($\alpha$)\\
11& $epde$ & yr & Mean Epoch in Dec. ($\delta$)\\
13 & $jmag$ & mag & 2MASS $J$ magnitude \\
14 & $e\_jmag$ & mag & error of 2MASS $J$ mag.\\
15 & $hmag$ & mag & 2MASS $H$ magnitude \\
16 & $e\_hmag$ & mag & error of 2MASS $H$ mag.\\
17 & $kmag$ & mag & 2MASS $K$ magnitude \\
18 & $e\_kmag$ & mag & error of 2MASS $K$ mag.\\
19 & $b1mag$ & mag & $B$ mag: USNO-B, 1st epoch \\
20 & $b2mag$ & mag & $B$ mag: USNO-B, 2nd epoch \\
21 & $r1mag$ & mag & $R$ mag: USNO-B, 1st epoch \\
22 & $r2mag$ & mag & $R$ mag: USNO-B, 2nd epoch \\
23 & $imag$ & mag & $I$ mag: USNO-B \\
24 & $surveys$  & -- & Origin of USNO-B mags\\
25 & $nobs$ & -- & total number of astrometric observations ($n_{\rm PPMXL}+1$)\\
26 & $gaia\-id$ &--& Gaia unique source identifier\\
27 & $Gmag$& mag & mean Gaia $G$-band magnitude\\
28 & $e\_Gmag$& mag & estimated error of Gaia $G$-mag\\
29 & $clone$ & -- & $>$1 PPMXL match to Gaia object \\
30 & $no\_sc$ &-- & object not in SuperCOSMOS \\
\hline                                             
\end{tabular}
\end{table}
In the meantime, we present a very short-lived yet powerful astrometric catalog, adding value and scientific use cases to Gaia DR1. It provides proper motions
for 583 million objects, i.e. for more than half of the objects
 for which Gaia DR1 only gives positions. This is achieved by combining data from the PPMXL catalog \citep{2010AJ....139.2440R} and Gaia DR1 positions. Named ''HSOY'' ({\bf H}ot {\bf S}tuff for {\bf O}ne {\bf Y}ear), highlighting
its short-lived nature, we intend to partly fill the gap in time between the ultra-precise positions of DR1 and the 
ultra-precise full 5 parameter astrometry of DR2. 

Until HSOY will be superseded by Gaia DR2, it presents the best set of proper motions in existence in the magnitude range fainter than TGAS to $G=20$~mag,
and is a valuable base for studies of stellar kinematics. 

Sect. \ref{sect:constr} describes the assembly of this catalog and its input data, as well as giving the overall characteristics of this catalog.
In Sect. \ref{sect:science} 
we demonstrate the improvement of the precision of proper motions of HSOY with respect to current entirely ground-based values with two science case examples.
\section{Presenting  HSOY}
\label{sect:constr}
\subsection{Construction and stellar content of  HSOY} \label{sect21}
HSOY has been constructed using the method previously used to assemble the PPMXL catalog \citep{2010AJ....139.2440R}, 
which itself now forms one of the input datasets for HSOY.
 For PPMXL the input data were the  
2MASS \citep{2006AJ....131.1163S} and USNO-B1.0 catalogs \citep{2003AJ....125..984M}, for HSOY, accordingly, PPMXL and Gaia DR1. The procedure of construction,
 described in detail
in \citet{2010AJ....139.2440R}, involves cross-matches between the datasets, and a weighted least-squares fit to derive positions and proper motions. 


PPMXL contains about 900 million, and Gaia DR1 1.1 billion sources. Yet HSOY 
only contains 583,001,653 entries, i.e. about 50-60\% of the object 
numbers of the input catalogs. Fig. \ref{hsoy_logn.fig} shows the object counts for both
HSOY and Gaia DR1. 
Of course, HSOY
can only contain objects which are in both PPMXL and
Gaia DR1. Objects that did not make it into the final HSOY are
very probably non-stellar objects and failed matches
originating in the USNO-B1. However,
the inhomogeneous sky coverage of Gaia DR1 \citep{Gaia2} most likely also plays a role.

On the other hand, there
  still is a significant fraction of entries probably not related to
  physical objects in HSOY.  The most common form of these are spurious pairs. 
These may arise from observations that have not been
  matched in USNO-B, either from different epochs or from different
  plates.  As long as the original USNO-B matched up two observations
  of the same object for each pair member and the observations had
  sufficient precision, they will form a close, common proper motion
  pair in PPMXL and will consequently be matched to the same
  Gaia DR1 object.  Such objects (and a few other cases where two or
  more PPMXL objects are matched to the same Gaia DR1 object) are
  marked with a non-NULL clone flag (0.7\% of the entire catalog).
  PPMXL contained about 24.5 million objects with proper motions larger than 150 mas/yr
on the northern hemisphere, alone against an expectation
  of about $10^5$ as discussed in the PPMXL paper.  The procedure
  outlined above brings the number of high-PM objects in HSOY down to 2.5
  million on the entire sky ($2.5\cdot 10^5$ in the north).  Hence, there
  are still many spurious high PM objects in HSOY accidentally matching a
  (real) Gaia DR1 object at J2015.  Another reduction of the
  spurious sources can be effected by matching PPMXL against
  SuperCOSMOS \citep{2001MNRAS.326.1279H}, an independent source extraction from the plate
  collections underlying USNO-B at J2000.  Where no such match can be
  found within $3''$, HSOY set the \textit{no\_sc} column to 1. On
  the northern sky, only using objects with matches in SuperCOSMOS, only
  168206 objects with $\mu>150\,{\rm mas/yr}$ are left, within a factor
  of two of the level to be expected from LPSM
  \citep{2005AJ....129.1483L}.  All-sky, including the very crowded
  fields on the southern sky, there are about $1.38\cdot 10^6$ high-PM
  objects with matches in SuperCOSMOS in HSOY.

  Conversely, sometimes more than one Gaia DR1 object is within one
  PPMXL objects' match radius.  While in come cases, this may be due to
  true binaries already resolved by Gaia, more typically they will be
  due to failed observation matching in the construction of Gaia DR1
  and should therefore generally be considered spurious pairs, too.
  They are marked with a non-NULL comp flag (1.5\% of the entire
  catalog).
In both catalogs, there are a couple of hundred sources fainter than 21 mag, see Fig. \ref{hsoy_logn.fig}. 
These have to be considered spurious sources.
\subsection{The astrometric precision of  HSOY} \label{sect22}
For the positions, the 
overwhelming precision of Gaia DR1 results in mean epochs close to that of Gaia DR1 of 2015.0; the mean epoch of most objects in HSOY is near 2014.8. 
In HSOY, the positions are given for epoch J2000.0 by applying
proper motions. Also, the formal precision of these positions is entirely determined by
the precision of the proper motions. These are at maximum 5 mas/yr (see below), so the
positional rms-errors at J2000.0 are well below 0.1 arcsec,
and are not individually given in the catalog.
\begin{figure*}
   \centering
   \includegraphics[width=5.1cm]{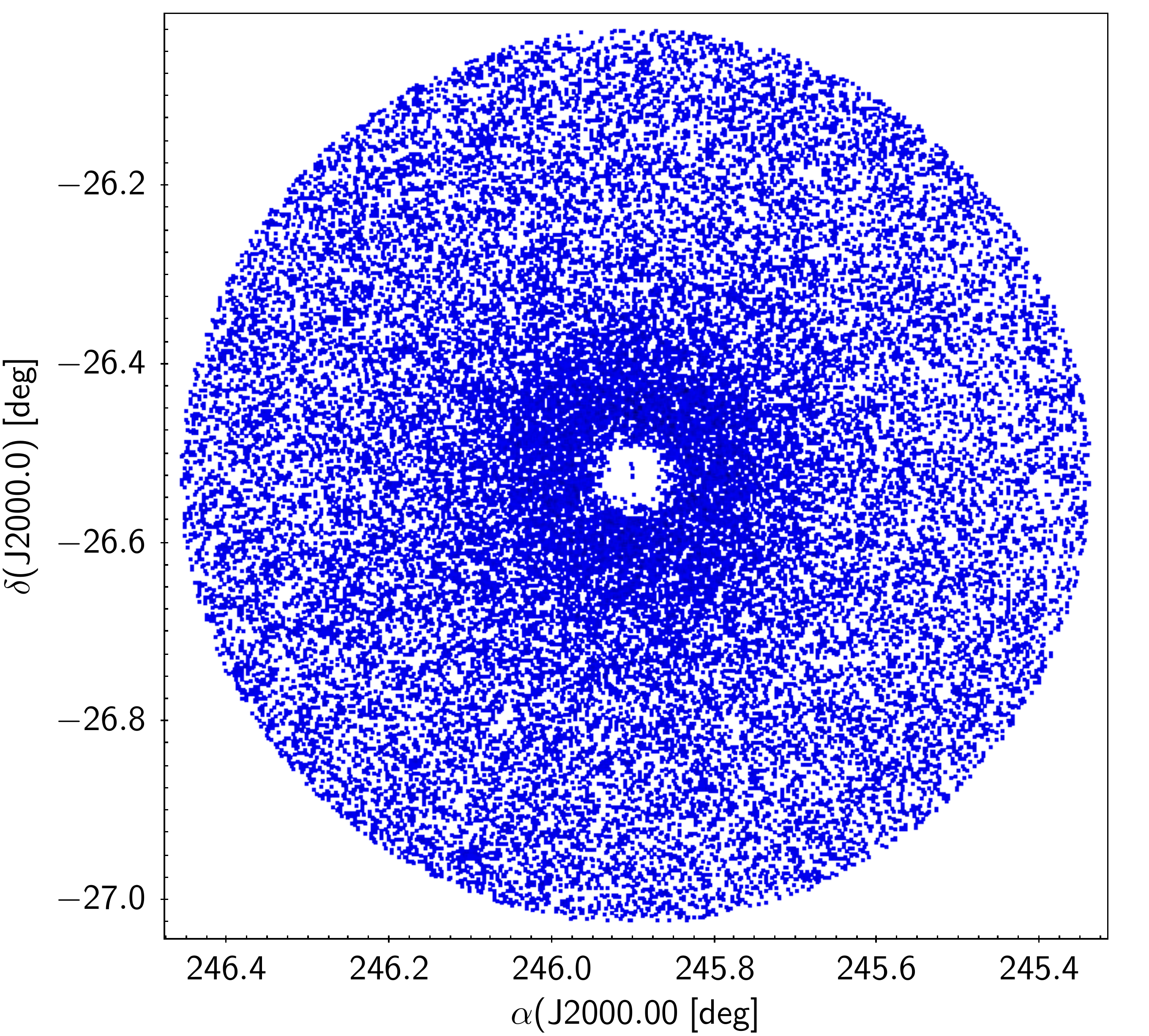}
   \includegraphics[width=6.0cm]{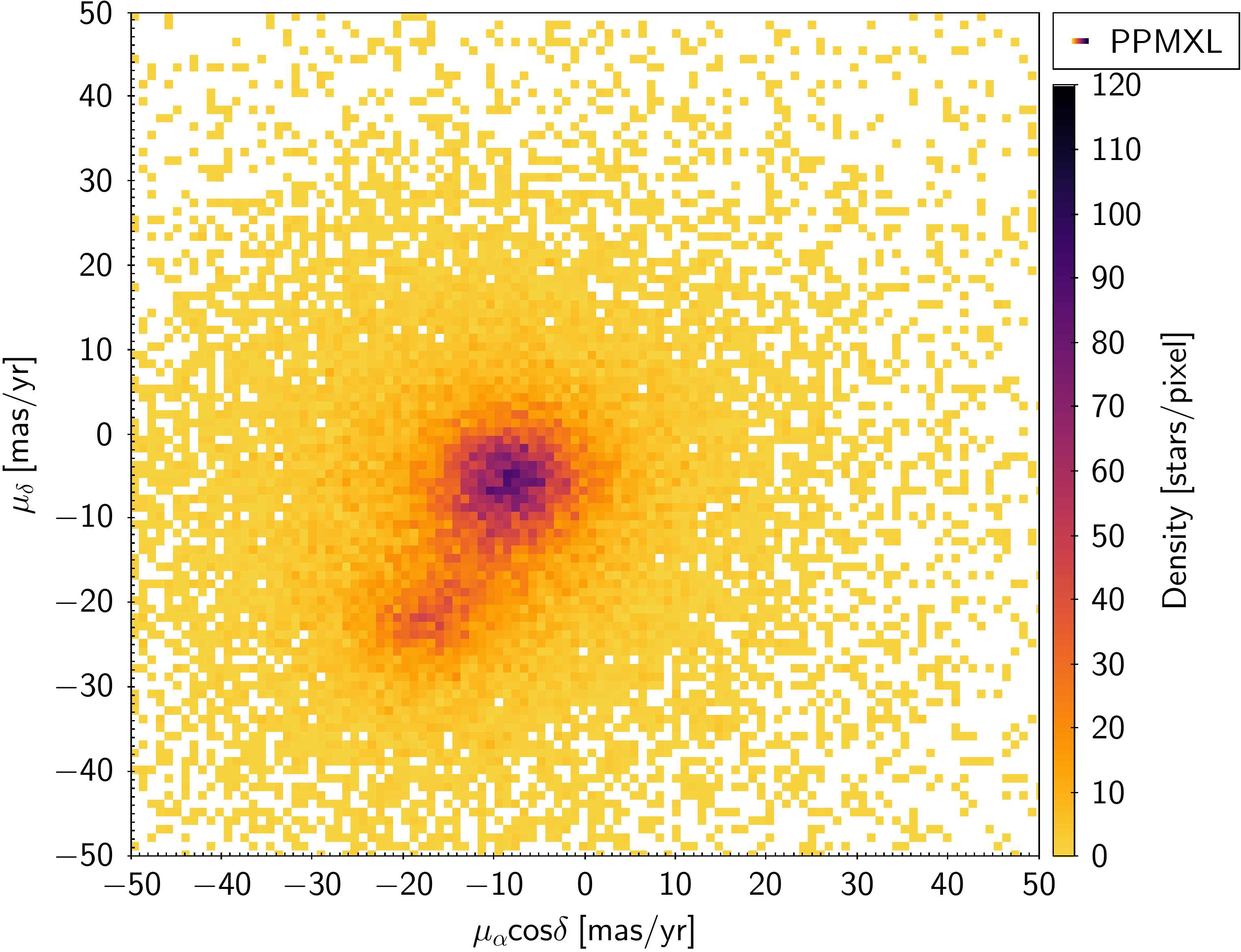}
   \includegraphics[width=6.0cm]{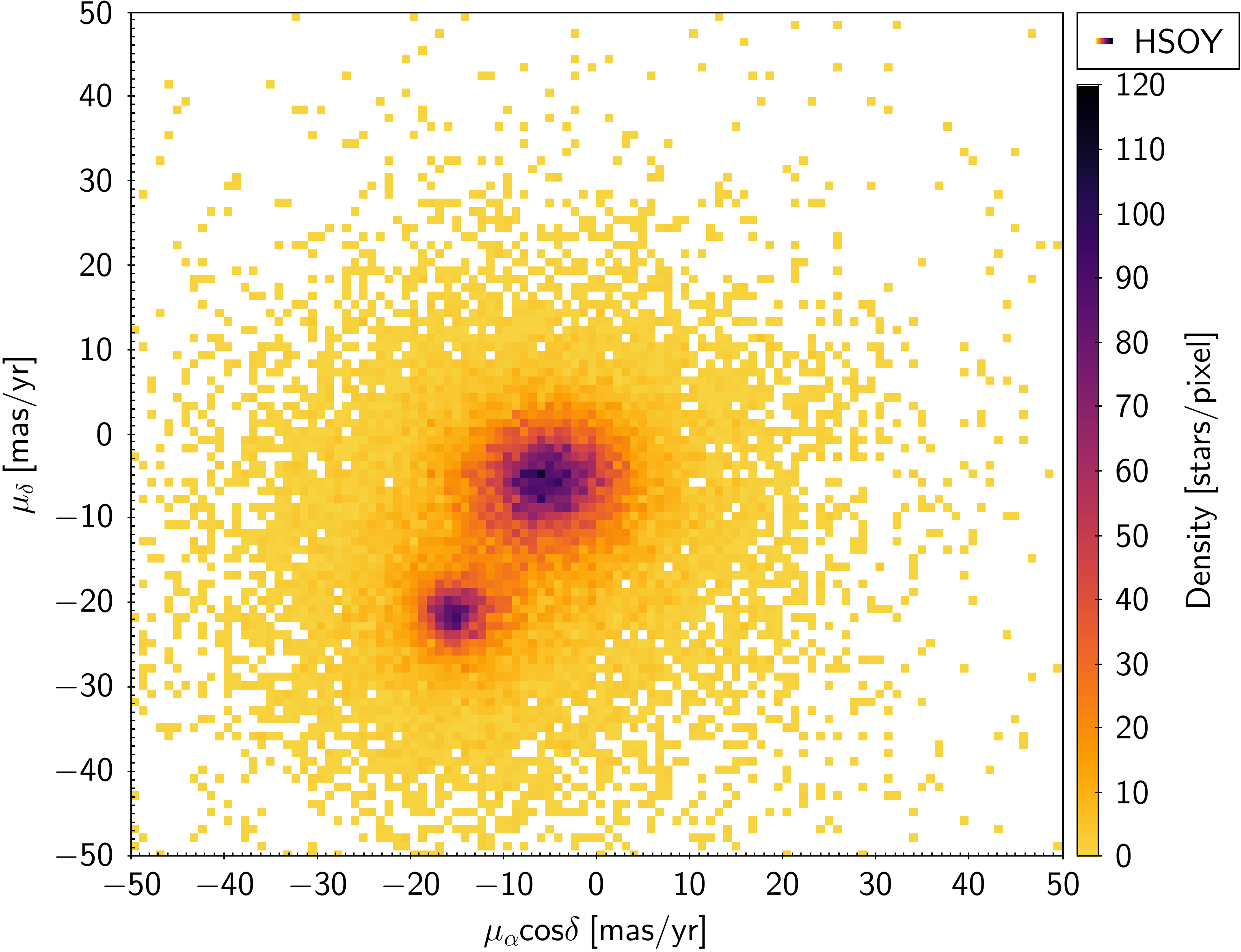}
   \caption{left panel: plot of the field with a radius of 30' taken from the HSOY catalog centered around M~4 .
    The center and right panels show the  vector point diagrams of the M~4 region, with proper motions taken from the PPMXL (centre) and HSOY (right). The hole in the middle
of the plot is caused by the strong crowding in the central part of M~4, the few points inside this hole can be considered as being spurious, which means they have to by
suppressed in any kind of analysis.}
              \label{m4.fig}%
    \end{figure*}
\begin{figure*}
   \centering
   \includegraphics[width=5.1cm]{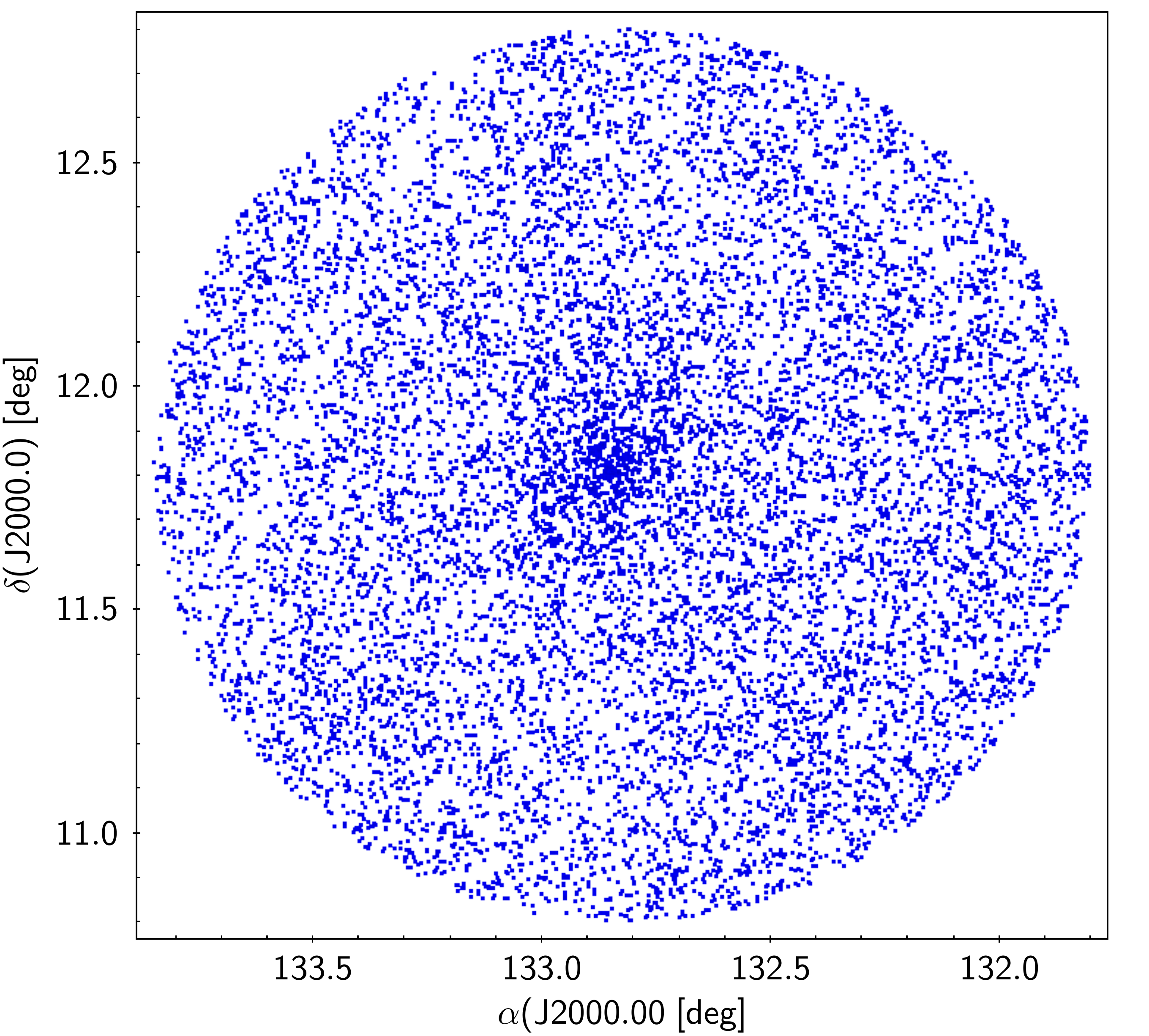}
   \includegraphics[width=6.0cm]{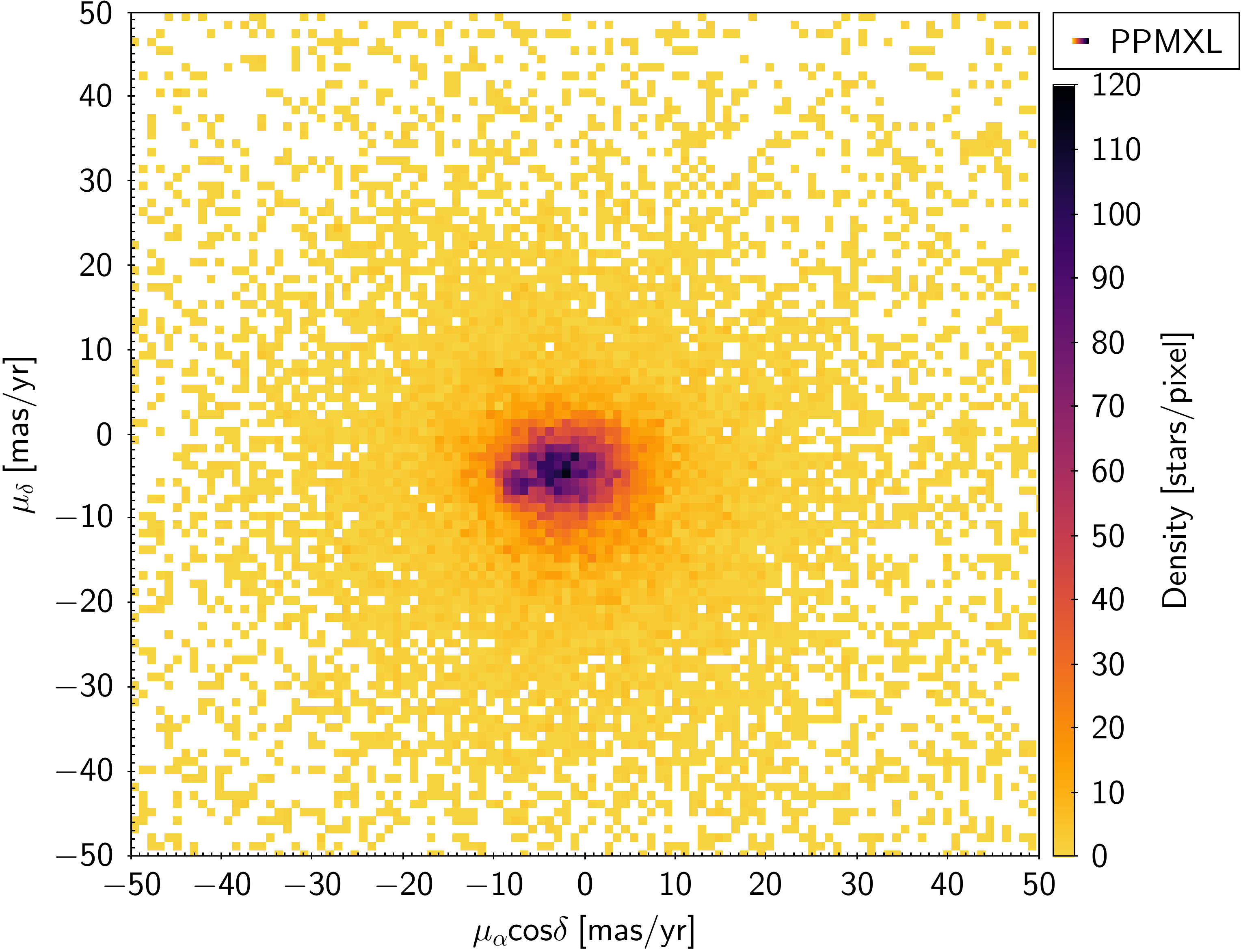}
   \includegraphics[width=6.0cm]{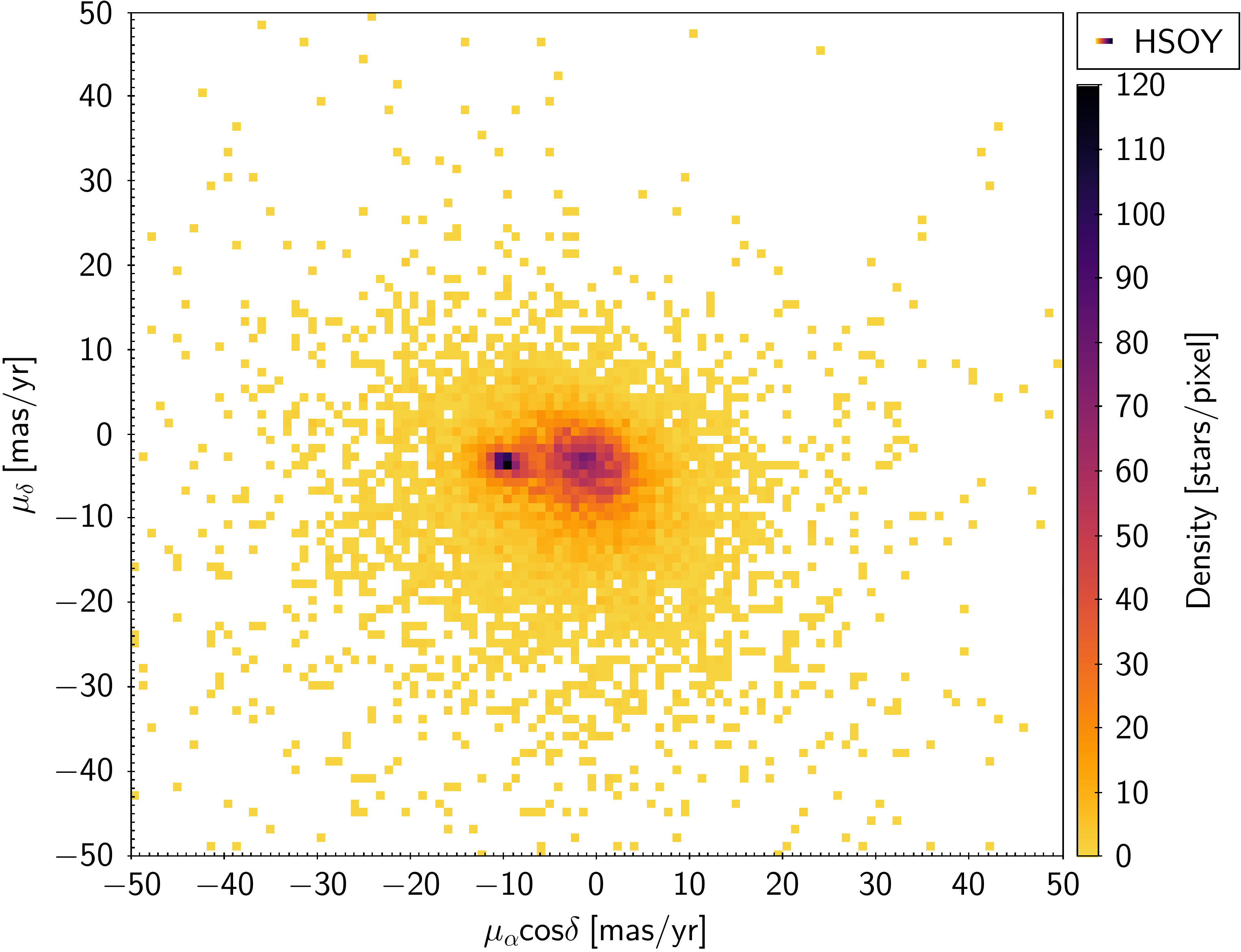}
   \caption{left panel: plot of the field with a radius of 1 degrees around M~67 taken from the HSOY catalog.
    The center and right panels show the  vector point diagrams of the M~67 region, with proper motions taken from the PPMXL (centre) and HSOY (right).
}
              \label{m67.fig}%
    \end{figure*}
Since for the HSOY-proper motions, PPMXL's positions are necessary, the vastly higher precision of the Gaia DR1-positions does not 
dominate as in the case of the mean-epoch positions in the HSOY catalog. 
This also means that they reflect some of the systematic errors in the PPMXL, such as the 
zonal errors present in all ground-based catalogs of similar type \citep{2010AJ....139.2440R}. This is to be kept in mind, when using HSOY proper motions. 
Due to the addition of Gaia DR1 these systematics
are reduced to a certain extent, but do not 
vanish altogether\footnote{It would in principle be possible to reduce these errors even further by re-constructing the PPMXL itself using Gaia DR1, 
but this is not worthwhile on the timescale of the useful life of HSOY}. Given the much smaller positional errors of Gaia DR1, the
correlations between the errors in RA and declination in Gaia DR1 can here be neglected. 
Generally the average formal errors for HSOY proper motions range from $<0.2$ mas/yr for stars brighter
than 8 mag, up to 4 or 5 mas/yr near the faint magnitude limit, see Fig. \ref{hsoy_pmerrs.fig}. 

Fig. \ref{hsoy_global.fig} exhibits much larger errors in both proper motion components at declinations south
of $-30$\degr  - the mean formal errors
there are slightly less than double of those in the rest of the sky, as shown in Fig. \ref{hsoy_pmerrs.fig}.
This is inherited from the underlying plate surveys: the first all-sky Schmidt plate
surveys started in
the northern hemisphere in the 1950s and were extended to the
south only much later in the 1970s. Therefore the baseline for
proper motions in the southern quarter of the sky is shorter by
20 years, with the corresponding consequences for the formal
proper motion uncertainties.
Apart
from this issue the errors in proper motion over the whole sky are remarkably
homogeneous, being just a little higher near the dense areas of
the Milky Way. Note that we used the original PPMXL proper motions rather than the
possibly more inertial ones given by \citep{2016AJ....151...99V} since the
latter are only available for objects with 2MASS photometry and hence
less than half the PPMXL.


HSOY is not a dedicated photometric catalog; therefore it utilises all photometry that its input catalogs supply. From its PPMXL parent, it retains the photographic
magnitudes taken from the USNO-B1 catalog and the NIR 2MASS values. Added to this is the Gaia DR1 $G$-magnitude. Therefore, for more information regarding
 the quality of the photometry, we refer to the
original sources, i.e. Gaia DR1, USNO-B1, and 2MASS.

As for Gaia DR1 itself, the primary access mode to HSOY is the
  Virtual Observatory protocol TAP 
through the service at http://dc.g-vo.org/tap.  Further
  access options are discussed at http:///dc.g-vo.org/hsoy
  \citep{vo:hsoy_main}. Table \ref{tab1.tab} shows the data content of HSOY.

\label{sect:char}
\section{The proper motions of M~4 and M~67 as science case examples}
\label{sect:science} 
In order to demonstrate the increased capabilities of HSOY, especially with respect to earlier, entirely ground-based catalogs, e.g. PPMXL, we present the proper motion 
distributions in the fields of the globular cluster M~4 (NGC 6121) and the rich, old open cluster M~67 (NGC 2682). These objects are especially instructive, given their
 large proper motion and high stellar density. 

For M~4 we downloaded a circular field with a radius of 0.5\degr, which was found to show both the cluster and the field stars best, from both catalogs. The field of
view is shown in Fig.~\ref{m4.fig}. The resulting  vector point diagrams (VPD) are also shown in Fig.~\ref{m4.fig}. A comparison of the VPD made from the PPMXL proper motions 
(centre panel) with that made from HSOY data clearly shows the dramatic improvement. Although the PPMXL has significantly more stars in this field than HSOY (42,000 vs. ~30,000), the M~4 proper motion peak is much more and the field peak somewhat more pronounced in the latter case.

The open cluster, M~67, one of the oldest of its kind and rather populous, is our second demonstration object. Fig.~\ref{m67.fig} shows the field and the  vector point diagrams
for both PPMXL (centre) and HSOY (right), this time, given the less dense environment, using a field with a radius of 1\degr. Again the clear improvement is seen by comparing the
centre and right panels in Fig. \ref{m67.fig}. While the PPMXL only shows hints of the cluster, it clearly stands out in the VPD generated using HSOY.
 

Both demonstration cases, i.e. M~4 and M~67 highlight the scientific potential of this catalog. The improvement is clearly shown.
There are certainly many other science cases which will profit from the 
existence of HSOY. 
\section{Outlook}
In a little more than a year from now, Gaia DR2 will be released and will thus make HSOY obsolete. However, we believe that this catalog will be put to good use until then.
In a way it is the final version of the second-generation ground-based astrometric catalogs, i.e., those done before or at the beginning of the CCD age, but with old photographic plates as the
long time-baseline basis. On the other hand it presents a bridge to a new generation of ground-based astrometric surveys, now based on Gaia data, such as what will
come out of LSST. These will go much fainter than either the current catalogs or Gaia, 
and will continue the tradition of space-calibrated ground-based astrometric 
catalogs. 
\begin{acknowledgements}
S. Roeser and E. Schilbach were supported  by  Sonderforschungsbereich
SFB  881  “The  Milky  Way  System”  (subprojects  B5)  of  the  German
Research Foundation (DFG). It is a great pleasure to acknowledge Mark Taylor from the Astrophysics Group of the School of Physics at the University of
Bristol for his wonderful work on TOPCAT, Tool for OPerations on Catalogues
And Tables and STILTS,  Starlink Tables Infrastructure Library Tool Set.
This research has made use of the 
resources of CDS, Strasbourg, France. Technical and
publication support was provided by GAVO under BMBF grant 05A14VHA.
This work has made use of data from the European Space Agency (ESA)
mission {\it Gaia} (\url{http://www.cosmos.esa.int/gaia}), processed by
the {\it Gaia} Data Processing and Analysis Consortium (DPAC,
\url{http://www.cosmos.esa.int/web/gaia/dpac/consortium}). Funding
for the DPAC has been provided by national institutions, in particular
the institutions participating in the {\it Gaia} Multilateral Agreement.

\end{acknowledgements}

\end{document}